# Self-consistent treatment of electrostatics in molecular DNA braiding through external forces

Dominic J. (O') Lee[1,a.)]

[1]Department of Chemistry, Imperial College London, SW7 2AZ, London, UK

**Abstract**

In this paper we consider a physical system in which two DNA molecules braid about each other. The distance between the two molecular ends, on either side of the braid, is held at a distance much larger than supercoiling radius of the braid. The system is subjected to an external pulling force, and a moment that induces the braiding. In a model, developed for understanding such a system, we assume that each molecule can be divided into a braided and unbraided section. We also suppose that the DNA is nicked so that there is no constraint of the individual linking numbers of the molecules. Included in the model are steric and electrostatic interactions, thermal fluctuations of the braided and unbraided sections of the molecule, as well as the constraint on the braid linking (catenation) number. We compare two approximations used in estimating the free energy of the braided section. One is where the amplitude of undulations of one molecule with respect to the other is determined only by steric interactions. The other is a self-consistent determination of the mean squared amplitude of these undulations. In this second approximation electrostatics should play an important role in determining this quantity, as suggested by physical arguments. We see that if the electrostatic interaction is sufficiently large there are indeed notable differences between the two approximations. We go on to test the self-consistent approximation, included in the full model, against experimental data, for such a system, and we find good agreement. However, there seems to by a slight left/right handed braid asymmetry in some of the experimental results. We discuss what might be the origin of this small asymmetry

## 1. Introduction

In biological systems DNA braiding occurs and has an important role to play. Two notable examples of DNA braiding are the formation of plectonemes and dual molecule catenanes. The former are a natural state for plasmids in bacteria [1], and plectonemes are also formed in the replication process [2] and in transcription [3] by the unravelling of the DNA strands. Catenane structures are seen between the two daughter DNA molecules as intermediates in the replication of a DNA plasmid [4,5].

In the past, to try understand DNA braiding in plectonemes, single molecule twisting experiments have been performed [6,7,8,9,10,11,12,13]. However, only recently, has there also been an interest in performing dual braiding experiments [14,15,16], which may provide insights into braiding in DNA catenane like structures. The experiments, reported in Refs. [14] and [16], involve the two DNA molecules being attached by antibodies to a substrate and a magnetic bead. The magnetic field that is applied to the bead provides a pulling force that stretches out the DNA molecules, supressing the undulations of the molecular centrelines, and a moment that produces a fixed number of turns of the bead, and so a braid.

---

[a.)] Electronic Mail: domolee@hotmail.com

To describe single molecule twisting experiments and plectoneme supercoiling a considerable amount of theoretical work has been performed [17,18,19,20,21,22,23,24,25,26,27, 28,29,30]. However, little work has been done on the statistical mechanics of dual molecular braiding [31,32]. However, in two recent publications [33, 34], we have developed a new model that describing braiding experiments of the form of those done in Ref. [16]. These models allow for the two ends of the two DNA molecules to be apart a distance much larger than diameter of the braid which they form, which is relevant for the experiments of Ref. [16], as the DNA are attached to the bead with an end to end separation between the two molecules of the order of one micron. Also, one important improvement over the work of Ref. [32], is that we allow for a mean-field braid structure, which is self-consistently determined, that allows us to go to larger braid linking (catenation) numbers. We now allow for electrostatic interactions between molecules.

In the models of Refs. [33] and [34], the size of undulations of molecules, relative to each other, within the braid is determined only by steric interactions between the two molecules. However, when other interactions- for instance electrostatic interactions- are present the size of these undulations should be self-consistently determined. There is a simple argument to suggest this. Firstly, the size of undulations affects the average strength of intermolecular interactions. Therefore, if these interactions are repulsive, it is energetically unfavourable for molecules to have large amplitude undulations relative to each other. Thus, to reduce the free energy, the size of undulations will be reduced when repulsive interactions are present. A self-consistent treatment of the amplitude undulations was originally pioneered in Ref. [24] for braids in plectonemes, and has been shown to fit the data of single molecule twisting experiments much better [28] than supposing that just steric interactions determine the mean squared amplitude of undulations. With this in mind, we wanted to introduce a self-consistent determination of the mean squared amplitude of such undulations into the theory. This treatment is along similar lines to Ref. [24], however steric forces between the two molecules are also taken into account.

The work is presented in the following way. In the next section, we start by reviewing general features of the model; starting with the generic form, we originally used in Refs. [33] and [34]. We then discuss the various contributions that we include in the free energy of the braided section of the two molecules. Next, a formula for the free energy of the braid is presented for the approximation used in Refs. [33,34]. Here, the mean squared amplitude of undulations, is determined only by steric interactions. We call this the simple approximation. Then, last of all, we present the form of the free energy when the mean squared amplitude of undulations is determined self consistently from both electrostatic and steric interactions. The results section is divided into two parts. In the first part we compare the self-consistent approximation with the simple approximation, for different strengths of the electrostatic interaction between molecules. We show results for two geometric parameters that characterize the average structure of the braid as a function of the number of induced braid pitches. Also, we show the applied moment required to generate a particular number of braid turns (pitches) and the end to end extension of the two molecules. We see, indeed, as the strength of the electrostatic interaction is increased the difference between the two approximations increases. In second part, we compare the self-consistent theory with experimental data of Ref. [16] and find good agreement with the model. In the last section, our discussion and outlook, we discuss extensions to the work, as well as the possibility that weak chiral interactions may account for the slight asymmetry between left and right handed braids seen in some of the experimental data.

## 2. Model

### *2.1 General considerations*

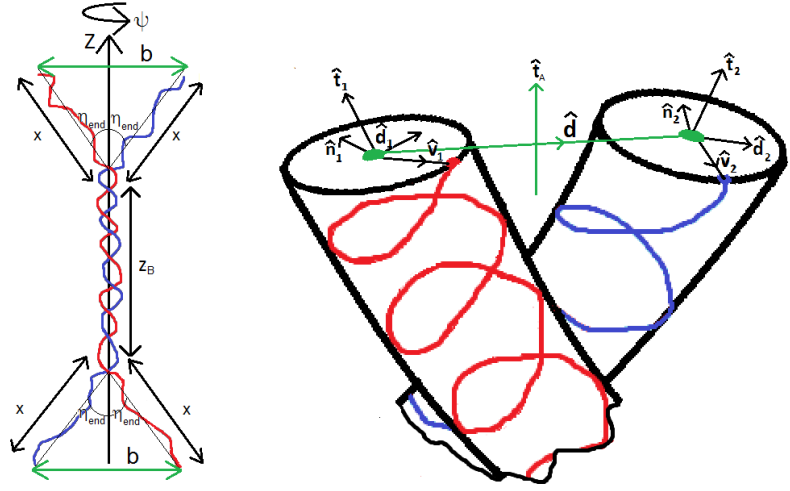

Fig.1 Schematic picture of the configuration of the two molecules. The left hand picture shows the global configuration of two molecules. The red (lighter) line denotes one molecule, while the blue (darker) line denotes the other one. Both sets of molecular ends are separated distance $b$ apart and one set of ends is rotated an angle $\psi$ with respect to the other set about the axis of the braid. The distance $x$ is the distance along the average position of the centre lines of the end segments and is given by the WLC formula $x = (L - L_b)(1 - \sqrt{k_B T \cos(\eta_{end}/2)/2l_p F})$ (see Refs. [33] and [34]), where $l_p$ is the bending persistence length of the two molecules. To generate this configuration, two of the ends may be attached to a magnetic bead and the other two ends to a substrate, as in the experiments of Ref. [16]. In the second picture we show the tangent vectors ($\hat{\mathbf{t}}_1(\tau)$ and $\hat{\mathbf{t}}_2(\tau)$) of the two molecular centre lines, the vector $\hat{\mathbf{d}}(\tau)$ that lies along a line connecting the two molecular centre lines (shown in green) and the tangent vector of the braid centre line that define the local configuration of the braid. Also shown are the vectors $\hat{\mathbf{n}}_\mu(\tau)$ and $\hat{\mathbf{d}}_\mu(\tau)$ ($\mu=1,2$), which are defined in Eqs. (8) and (9), that define the braid frames [30] of the two molecules. When $R'(\tau) = 0$ we have that $\hat{\mathbf{d}}_1(\tau) = \hat{\mathbf{d}}_2(\tau) = \hat{\mathbf{d}}(\tau)$. The vectors $\hat{\mathbf{v}}_\mu(s)$ characterize the azimuthal orientation of the minor grooves of both molecules, the trajectories of which are shown by the two distorted helices in the right hand side picture.

In this paper we use a model, developed in Refs. [33] and [34], that describes braiding of two DNA molecules of identical length $L$, the value of which is assumed large enough for finite size effects not to be important. The two sets of molecular ends are held apart by distance $b$ (see Fig. 1). One set of ends remain fixed, while the other set of ends are free to rotate about a common axis, which is assumed to be the axis of the braid (for a definition see below). To this system a pulling force $F$, in the direction along the braid axis, and a moment $M$ that rotates the molecular ends about the same axis, are applied. In the model, we divide the DNA molecules into unbraided end pieces and a braided central section. This allows us to write down the following free energy for our system

$$\mathcal{F}_T = 2(L - L_b) f_{wlc} + L_b f_{Braid}, \tag{1}$$

where $2(L-L_b)f_{wlc}$ is the contribution from the end pieces and $L_b f_{Braid}$ is the contribution from the braided section. $f_{Braid}$ will be taken to be a function of both $M$ and $F$. As before [33,34] the four end pieces are assumed behave like wormlike chains of contour length $(L-L_b)/2$, where $L_b$ is the contour length of each of the two molecules that contributes to the braid.

Through the WLC model [35], it is possible to relate $L_b$ to the parameters $b$, $L$, $F$ and $\frac{\eta_{end}}{2}$ [33,34] ; the last is the angle both average molecular centre lines, of the end sections, make with the axis of the braid (see Fig. 1). This relationship reads as

$$L_b \approx \left[L-\left(1+\left(\frac{k_B T \cos\left(\frac{\eta_{end}}{2}\right)}{2F l_p}\right)^{1/2}\right)\frac{b}{\left|\sin\left(\frac{\eta_{end}}{2}\right)\right|}\right]\theta\left(L-\left(1+\left(\frac{k_B T \cos\left(\frac{\eta_{end}}{2}\right)}{2F l_p}\right)^{1/2}\right)\frac{b}{|\sin\left(\frac{\eta_{end}}{2}\right)|}\right),$$

(2)

where $\theta(y)$ is the theta function, which for $y \geq 0$ is 1, otherwise zero, preventing a negative unphysical value of $L_b$. Here, $l_p$ is the bending persistence length of the DNA molecules. For DNA we take the value $l_p \approx 500 \text{Å}$.

Followng Refs. [33] and [34], the free energy density of the end pieces is found to be

$$f_{wlc}(b,F)(L-L_b) \approx -\frac{Fb}{2\left|\sin\left(\frac{\eta_{end}}{2}\right)\right|}\left(\cos\left(\frac{\eta_{end}}{2}\right)-\left(\frac{k_B T}{2F l_p \cos\left(\frac{\eta_{end}}{2}\right)}\right)^{1/2}\right). \quad (3)$$

The total end to end distance of the two molecules is given by

$$z_T \approx \langle z_B \rangle + \sqrt{(L-L_b)^2\left(1-\sqrt{\frac{k_B T \cos(\eta_{end}/2)}{F l_p}}\right)^2 - b^2}, \quad (4)$$

where the average length of the braid axis $\langle z_B \rangle$ (angular brackets will always correspond to thermal averaging) is determined from

$$\langle z_B \rangle = -L_b \frac{\partial f_{Braid}(M,F)}{\partial F}. \quad (5)$$

It is useful to define the number of braid turns, $N$ as the (average) number of times the two molecules wrap around each other (number of pitches) in the braided section. This quantity is related to $n$ through the relation

$$n \approx N + \frac{\text{sgn}(n)}{2}, \quad (6)$$

where $N$ is determined from

$$N = \frac{L_b}{2\pi} \frac{\partial f_{Braid}}{\partial M}. \quad (7)$$

*2.2 Simple Statistical Mechanical Model describing the braided section*

For the thermally averaged structure of the braid, we assume that the two molecular centre lines precess at a constant, and at the same, spatial frequency $\omega_{b,0}$ around a common axis that lies along the centre of the braid; this axis is what we define as the braid axis. It is along and about this axis $F$ and $M$ act, respectively. Equivalent positions along the two molecular centre lines may be described by an arc-length coordinate $s$ that runs from $-L_b/2$ to $L_b/2$. At any point along the braid, we may construct a line of length $R$ that connects two points on the molecular centre lines with the same arc-length coordinate $s$. For the thermally averaged braid structure, $R(s) = R_0$ and is constant with respect to $s$. Indeed, the braid axis bisects the midpoint of this line. Pointing along this line is the unit vector $\hat{\mathbf{d}}(s)$, from which we may define unit vectors $\hat{\mathbf{d}}_1(s)$, $\hat{\mathbf{d}}_2(s)$, $\hat{\mathbf{n}}_1(s)$ and $\hat{\mathbf{n}}_2(s)$

$$\hat{\mathbf{n}}_1(s) = \frac{\hat{\mathbf{t}}_1(s) \times \hat{\mathbf{d}}(s)}{|\hat{\mathbf{t}}_1(s) \times \hat{\mathbf{d}}(s)|}, \qquad \hat{\mathbf{d}}_1(s) = \hat{\mathbf{n}}_1(s) \times \hat{\mathbf{t}}_1(s), \quad (8)$$

$$\hat{\mathbf{n}}_2(s) = \frac{\hat{\mathbf{t}}_2(s) \times \hat{\mathbf{d}}(s)}{|\hat{\mathbf{t}}_2(s) \times \hat{\mathbf{d}}(s)|}, \qquad \hat{\mathbf{d}}_2(s) = \hat{\mathbf{n}}_2(s) \times \hat{\mathbf{t}}_2(s). \quad (9)$$

These vectors (shown in Fig. 1) define a 'braid frame' [30], which can be used to characterize the local azimuthal orientations of the cross-sections of the two molecules.

Also, to characterize the geometry of the braided section, we define a braid tilt angle $\eta(s)$. This is the angle between the two tangent vectors of the molecular centre lines, $\hat{\mathbf{t}}_1(s)$ and $\hat{\mathbf{t}}_2(s)$ respectively, such that

$$\hat{\mathbf{t}}_1(s).\hat{\mathbf{t}}_2(s) = \cos\eta(s). \quad (10)$$

In the thermally averaged braid structure $\eta(s) = \eta_0$. The angle $\eta_0$ is assumed constant with respect to $s$, as it is related to the average frequency of precession $\omega_{b,0}$ of the two molecular centre lines and $R_0$, which are constant. We allow for the braid to thermally fluctuate about this average structure by allowing for fluctuations in both $\eta$ and $R$ about the mean field values $\eta_0$ and $R_0$. We also allow for small thermal undulations in the braid axis away from its average configuration of a straight line. All of these fluctuations in the local geometry depend on $s$. A particular configuration of the braid is assigned a Boltzmann weight in the partition function depending on its total energy. There are four contributing factors to the energy that we take account of, and thus in the free energy density $f_{Braid}$.

The first is the bending elastic energy of the two molecules forming the braid, which is described by the elastic rod model for DNA. Here, the elastic energy is computed by integrating the sum of the squares of the curvatures, for both molecular centre lines, along the lengths of the molecules contributing to the braided section and multiplying by $k_B T l_p / 2$ (for expressions for the elastic energy contribution see Refs. [33] and [34]). The second contribution is an electrostatic energy between helices [36] of the form

$$E_{int} = \int_{-L_b/2}^{L_b/2} ds \left( \mathcal{E}_{dir}(R(s)) + \mathcal{E}_{img}(R(s)) \right). \tag{11}$$

Here, $\mathcal{E}_{dir}(R(s))$ is the contribution from direct electrostatic interactions between the charges of one molecule and the other, and is given by

$$\frac{\mathcal{E}_{dir}(R)}{k_B T} = \frac{2 l_B (1-\theta)^2}{l_c^2} \frac{K_0(R\kappa_0)}{\left[ a\kappa_0 K_1(a\kappa_0) \right]^2}. \tag{12}$$

The term $\mathcal{E}_{img}(R(s))$ is also repulsive, having effectively half the decay range. It is the contribution due to one molecule interacting with its image charge reflection at the surface of the other molecule (see Ref. [36]). It takes the form

$$\frac{\mathcal{E}_{img}(R)}{k_B T} = -\frac{2 l_B}{l_e^2} \sum_{n=-\infty}^{\infty} \sum_{j=-\infty}^{\infty} \left( \delta_{n,0} \theta - \cos(n\phi_s) \right)^2 \frac{K_{n-j}(R\kappa_n) K_{n-j}(R\kappa_n)}{\left[ a\kappa_n K_n(a\kappa_n) \right]^2} \frac{I'_j(a\kappa_n)}{K'_j(a\kappa_n)}. \tag{13}$$

In both Eqs. (12) and (13) we have that

$$\kappa_n = \sqrt{\frac{1}{\lambda_D^2} + \left( \frac{2\pi n}{H} \right)^2}; \tag{14}$$

the Debye screening length, $\lambda_D$; the Bjerrum length, $l_B$ (taken to be $l_B \approx 7\text{Å}$); and the length $l_e \approx 1.7\text{Å}$, which is the inverse of the average DNA linear charge density multiplied by the unit charge $e$. The parameters $a$, $\phi_s$ and $H$ are the effective DNA radius (for electrostatics), the angular half width of the minor groove and the average value of the DNA pitch. We choose the values $a \approx 11.2\text{Å}$, $\phi_s \approx 0.4\pi$ and $H \approx 33.8\text{Å}$. Eqs. (11)-(13), are a simplification of those used in Refs. [37], [38] and [33], derived using the mean field electrostatic model of Ref. [36] (for the most general calculation of the electrostatic energy see Ref. [39]). We have supposed, in Eqs. (11) and (12), that any forces depending on helix structure, in $\mathcal{E}_{dir}(R)$, are completely washed out by thermal fluctuations (this corresponds to taking the limit $\lambda \to \infty$ in Eq. (9) of Ref. [33]). In this case, valid when helix specific forces are sufficiently weak, localizing ions near the DNA grooves has a small effect on the results. Furthermore, for monovalent salt ions (with the notable exception of some transition metals), we do not expect a large degree of localization [40]. Therefore, for simplicity, we suppose that the layer of condensed counterions is uniformly distributed near the DNA surface, compensating the bare DNA charge by a fraction $\theta$. We will see Eqs.(11)-(13) seem to adequately describe the electrostatic interaction in the monovalent salt experiments of Ref. [16]. Though, as a correction to Eq. (11), there may well be a weak residual chiral interaction from correlations between the helix structures of the two molecules that leads to a very slight $n \to -n$ asymmetry seen in some of the results of Ref. [16]. We will discuss this possibility later in the discussion section.

The third thing that we consider is steric interactions. Here, we assume that DNA molecules can be modelled as hard core cylinders with steric radius $a$ (see Refs. [38] and [41]). In treating the steric interaction, in the statistical mechanics, we use an approach originally developed in Ref. [42]. This is to replace the hard-core interaction with that of harmonic potential with an effective spring constant $k_{eff}$. This parameter depends on $R_0 - 2a$, which is a measure of the available space the molecules can fluctuate in, without colliding with each other. In the simple model, which we consider in this subsection, we assume the mean squared amplitude of undulations is primarily determined by steric interactions such that

$$\langle (R - R_0)^2 \rangle \approx (R_0 - 2a)^2. \tag{15}$$

The requirement, Eq. (15), allows us to determine an approximate relationship between $k_{eff}$ and $R_0 - 2a$ (see Refs. [38], [34] and [41] for details) [43].

The last contribution is a work term that contains both the moment $M$ and pulling force $F$ which is of the form

$$E_W = -2\pi \left( Tw_b + Wr_b \right) M - z_B F, \tag{16}$$

where applying the external moment $M$ changes the linking (catenation) number of the braid, $Lk_b = Tw_b + Wr_b$; the sum of braid twist and braid writhe, $Tw_b$ and $Wr_b$, respectively. In Eq. (16) we do not constrain the linking numbers of the individual molecules as in Ref. [32], therefore considering the DNA molecules as nicked. The braid twist may be defined as

$$Tw_b = \frac{1}{2\pi} \int_{-L_A/2}^{L_A/2} d\tau \omega_b(\tau), \tag{17}$$

where $\omega_b(\tau)$ is the spatial frequency of precession of the centre lines about the braid axis (note that $\langle \omega_b(\tau) \rangle = \omega_{b,0}$, and for an explicit expressions of $\omega_b(\tau)$ in terms of the braid geometric parameters see Ref. [41]). The coordinate $\tau$ is of unit arc length coordinate along the braid axis, runing from $-L_A/2$ to $L_A/2$ for the length of the braided section. The relationship between $L_A$ and $L_b$ depends on the configuration of the braid. In a configuration where the braid axis is straight $L_A = z_B$. The braid writhe is calculated through [44]

$$Wr_b = \frac{1}{4\pi} \int_{-L_A/2}^{L_A/2} d\tau \int_{-L_A/2}^{L_A/2} d\tau' \frac{(\mathbf{r}_A(\tau) - \mathbf{r}_A(\tau')) \cdot \hat{\mathbf{t}}_A(\tau) \times \hat{\mathbf{t}}_A(\tau')}{|\mathbf{r}_A(\tau) - \mathbf{r}_A(\tau')|^3}. \tag{18}$$

In Eq. (18), $\mathbf{r}_A(\tau)$ is the position vector that describes the trajectory of the braid axis and $\hat{\mathbf{t}}_A(\tau) = \frac{d\mathbf{r}_A(\tau)}{d\tau}$, the tangent vector. This decomposition of the braid linking number into braid twist and writhe for a molecular braid under tension was originally proposed in Ref. [32].

Strictly speaking, $Tw_b + Wr_b$, should be constrained to take exactly the value $-n + \text{sgn}(n)/2$ (when $b \to \infty$). However, instead of working in an ensemble where $Tw_b + Wr_b$ is exactly fixed, we work in a fixed $M$ ensemble, since it is much easier to do calculations. In the thermodynamic limit (the limit that we do our calculations in), where $L \to \infty$, these two ensembles are equivalent to each other. The fluctuations in the linking number in the fixed $M$ ensemble become negligible in this limit. As the braid axis is assumed to be straight, for the thermally averaged braid structure, we require $\langle Wr_b \rangle = 0$, (although for any given configuration in the thermal ensemble we may have $Wr_b \neq 0$). Thus, through Eqs. (7) and (16), we have in the thermodynamic limit $-N = \langle Tw_b \rangle$. If $M$ becomes sufficiently high, we may expect a buckled braided state where $\langle Wr_b \rangle \neq 0$, but we do not consider such a state in our present study. In Ref. [34] and [41] explicit expressions for both $z_B$ and $Tw_b$ are given, but for brevity we do not give them here. In calculating the free energy, the term $2\pi M Wr_b$ (in Eq. (16)) is considered to be small and is handled as a perturbation. This is done in similar way to the theoretical calculations of Ref. [22], describing single molecule twisting experiments. In the calculations of Ref. [22], the writhe and twist are to do with the molecular centre line and the trajectory of the minor groove (or some other point of reference that traces out the DNA double helix) about it.

Taking these contributions into account, following a variational approximation (the precise details of the calculation are given in Sections 1-8 of Ref. [41]), we obtain the following form for the free energy

$$\frac{f_{Braid}}{k_B T} = \left(\frac{F}{2l_p k_B T}\right)^{1/2} + \frac{\alpha_\eta^{1/2}}{2^{1/2} l_p^{1/2}} + \frac{1}{l_p^{1/3}} \frac{1}{2^{2/3}(R_0-2a)^{2/3}} - \frac{1}{R_0^2} \frac{l_p}{2^{1/3}} \left(\frac{R_0-2a}{l_p}\right)^{2/3} \sin^2\left(\frac{\eta_0}{2}\right)$$
$$+ \frac{4l_p}{R_0^2} \sin^4\left(\frac{\eta_0}{2}\right) + \mathcal{E}_{dir}(R_0) + \mathcal{E}_{img}(R_0) - \frac{F}{k_B T} \cos\left(\frac{\eta_0}{2}\right) + \frac{2M}{k_B T R_0} \sin\left(\frac{\eta_0}{2}\right) \quad (19)$$
$$- \frac{M}{4 k_B T R_0} \frac{1}{2^{1/3}} \left(\frac{R_0-2a}{l_p}\right)^{2/3} \sin\left(\frac{\eta_0}{2}\right)^{-1} - \frac{M^2}{16 l_p (k_B T)^2} \frac{1}{\cos\left(\frac{\eta_0}{2}\right)^4} \left(\frac{k_B T}{2 F l_p}\right)^{1/2},$$

where

$$\alpha_\eta = \frac{4l_p}{R_0^2} \left(3\cos^2\left(\frac{\eta_0}{2}\right)\sin^2\left(\frac{\eta_0}{2}\right) - \sin^4\left(\frac{\eta_0}{2}\right)\right) + \frac{F_R}{4 k_B T} \cos\left(\frac{\eta_0}{2}\right) - \frac{M_R}{2 R_0 k_B T} \sin\left(\frac{\eta_0}{2}\right). \quad (20)$$

In writing down Eq. (19), we have further assumed that the thermally averaged bending energy terms and electrostatic energy can be replaced with their unaveraged values at $R_0$. The first term in Eq. (19) is the free energy contribution due to undulations of the braid axis. The entropic contribution due to fluctuations in $\eta$ is given by the second term of Eq. (19). The third term is the contribution due to steric interactions between the molecules the braid and the entropy loss of confining the molecules in a braided configuration of radius $R_0$. The next two terms are contributions from the bending elastic energy of the molecules in the braid. Next in Eq. (19), we have the contribution to the free energy from the electrostatic terms, which is simply given by $\mathcal{E}_{dir}(R_0) + \mathcal{E}_{img}(R_0)$. The next three terms are the contributions from $-z_B F - 2\pi T w_b M$ in the work term described by Eq. (16). The last term is the leading order non-vanishing contribution in the perturbation series in $-2\pi W r_b M$ (see Ref. [41]). The $\cos(\eta_0/2)$ terms, in this last term, arise from the fact that unit arc-length of the braid axis, $\tau$ should be used in computing the braid writhe, not the arc-lengths of the molecular centre lines (see Eq. (18)). Exactly the same $M^2$ term (in the limit $\cos(\eta_0/2) \to 1$) was also computed in Ref. [32] in the case where the average linking number of each molecule is left unconstrained (note that in Ref. [32] one should set $F = 2f$). However, the whole approach goes beyond that of Ref. [32] in two regards. We include the confinement through steric interactions of $\delta R(s)$, fluctuations in the relative distance between the two centre lines, as well as considering fluctuations around a braided mean field configuration.

Equations that determine both $R_0$ and $\eta_0$ are then found through the minimization conditions

$$\frac{df_{Braid}}{dR_0} = 0 \quad \text{and} \quad \frac{df_{Braid}}{d\eta_0} = 0. \tag{21}$$

Also, both $\langle z_B \rangle$ and $N$ are related to $F$ and $M$ through Eqs. (5) and (7). By minimizing the total free energy, given by Eq. (1), with respect to $\eta_{end}$ we find that for sufficiently large pulling force

$$\cos\left(\frac{\eta_{end}}{2}\right) \approx -\frac{F}{f_{Braid}} - \left(\frac{2k_B T}{Fl_p}\right)^{1/2}\left(-\frac{F}{f_{Braid}}\right)^{3/2}. \tag{22}$$

***2.3 Self consistent determination of the mean squared amplitude of undulations of the braid***

In the case where there are just steric interactions, to maximise the entropy due to undulations, we simply have that $\langle (R-R_0)^2 \rangle \approx (R_0 - 2a)^2$. However, when we have repulsive electrostatic interactions, undulations enhance the strength of their thermal average, making large undulations energetically unfavourable. Therefore, the electrostatic interaction should also limit the size of $\langle (R-R_0)^2 \rangle$. Therefore, it seems that a better approach is to set $\langle (R-R_0)^2 \rangle \approx d_R^2$, where $d_R$ is self-consistently determined, primarily by electrostatic interactions, as well as steric forces. We also determine $\theta_R^2 = \left\langle \left(\frac{dR}{ds}\right)^2 \right\rangle$ self consistently.

In our expression for the free energy function we now use thermal averages of the bending energy and electrostatic energy terms; these averages help to determine $d_R$ and $\theta_R$. We do this according to a procedure used in Refs. [34] (for details see Ref. [41]), and [38]. The idea to introduce cut-offs on the amplitude of fluctuations in $R$, which we call $d_{min}$ and $d_{max}$. These cut-offs are the minimum and maximum values that $\delta R = R - R_0$ can take due to steric interactions. If $\delta R < d_{min}$ or $\delta R > d_{max}$, the values of both the bending and electrostatic energies are unphysical, as the molecules in the braid would have interpenetrated. Therefore, to prevent these unphysical values entering into the averaging, when $\delta R < d_{min}$ we replace $\delta R$ with $d_{min}$, and when $\delta R > d_{max}$ we replace $\delta R$ with $d_{max}$. We assume the values $d_{max} = -d_{min} = R_0 - 2a$, which should be adequate for the braids that we will study here, although a different choice might possibly be used for much more tightly wrapped braids (see Refs. [38] and [41]).

The parameters $d_R$ and $\theta_R$ are treated as a variational parameters that minimize the following free energy (which can be derived following the steps presented in Sections 10 -12 presented in Ref. [41])

$$\frac{f_{Braid}}{k_B T} = \left(\frac{F}{2l_p k_B T \cos(\eta_0/2)}\right)^{1/2} + \frac{\alpha_\eta^{1/2}}{2^{1/2} l_p^{1/2}} + \frac{d_R^2}{2^{8/3}(R_0-2a)^{8/3} l_p^{1/3}} + \frac{1}{4l_p \theta_R^2} + \frac{l_p \theta_R^4}{4d_R^2} - \frac{M^2}{16 l_p (k_B T)^2}\left(\frac{k_B T}{2l_p F_R}\right)^{1/2} \frac{1}{\cos\left(\frac{\eta_0}{2}\right)^{7/2}}$$

$$-\frac{L\theta_R^2 l_p}{R_0^2} \tilde{f}_1(R_0, d_R, R_0 - 2a, -(R_0 - 2a)) \sin^2\left(\frac{\eta_0}{2}\right) + \frac{2l_B(1-\theta_c)^2}{l_e^2 (a\kappa_D)^2 K_1(a\kappa_D)^2} g_0(\kappa_D R_0, \kappa_D d_R, (R_0-2a)/d_R, -(R_0-2a)/d_R)$$

$$-\frac{2l_B}{l_e^2} \sum_{n=-\infty}^{\infty} \frac{(\cos(n\phi_s) - \theta_c)^2}{(\kappa_n a K_n'(\kappa_n a))^2} g_{img}(n, \kappa_n R_0, \kappa_n d_R, (R_0-2a)/d_R, -(R_0-2a)/d_R; a)$$

$$+\frac{4l_p \tilde{f}_1(R_0, d_R, R_0 - 2a, -(R_0 - 2a))}{R_0^2} \sin^4\left(\frac{\eta_0}{2}\right) - \frac{F}{k_B T}\cos\left(\frac{\eta_0}{2}\right) + \frac{2M}{k_B T R_0}\sin\left(\frac{\eta_0}{2}\right) \tilde{f}_2(R_0, d_R, R_0 - 2a, -(R_0-2a))$$

$$-\frac{M\theta_R^2}{4k_B T} \frac{\tilde{f}_2(R_0, d_R, R_0 - 2a, -(R_0-2a))}{R_0} \sin\left(\frac{\eta_0}{2}\right)^{-1},$$

(23)

where now

$$\alpha_\eta = \frac{4l_p \tilde{f}_1(R_0, d_R, R_0 - 2a, -(R_0-2a))}{R_0^2}\left(3\cos^2\left(\frac{\eta_0}{2}\right)\sin^2\left(\frac{\eta_0}{2}\right) - \sin^4\left(\frac{\eta_0}{2}\right)\right) + \frac{F}{4k_B T}\cos\left(\frac{\eta_0}{2}\right)$$

$$-\frac{M\tilde{f}_2(R_0, d_R, R_0 - 2a, -(R_0-2a))}{2R_0 k_B T}\sin\left(\frac{\eta_0}{2}\right),$$

(24)

and

$$\tilde{f}_1(R_0, d_R, d_{\max}, d_{\min}) = \frac{R_0^2}{d_R \sqrt{2\pi}} \int_{d_{\min}}^{d_{\max}} \frac{dx}{(R_0+x)^2} \exp\left(-\frac{x^2}{2d_R^2}\right)$$
$$+\frac{1}{2}\left(\frac{R_0^2}{(R_0+d_{\min})^2}\left(1-\mathrm{erf}\left(-\frac{d_{\min}}{d_R\sqrt{2}}\right)\right) + \frac{R_0^2}{(R_0+d_{\max})}\left(1-\mathrm{erf}\left(\frac{d_{\max}}{d_R\sqrt{2}}\right)\right)\right),$$

(25)

$$\tilde{f}_2(R_0, d_R, d_{\max}, d_{\min}) = \frac{R_0}{d_R \sqrt{2\pi}} \int_{d_{\min}}^{d_{\max}} \frac{dx}{(R_0+x)} \exp\left(-\frac{x^2}{2d_R^2}\right)$$
$$+\frac{1}{2}\left(\frac{R_0}{(R_0+d_{\min})}\left(1-\mathrm{erf}\left(-\frac{d_{\min}}{d_R\sqrt{2}}\right)\right) + \frac{R_0}{(R_0+d_{\max})}\left(1-\mathrm{erf}\left(\frac{d_{\max}}{d_R\sqrt{2}}\right)\right)\right),$$

(26)

$$g_j(\kappa R_0, \kappa d_R, d_{\max}/d_R, d_{\min}/d_R) = \frac{1}{\sqrt{2\pi}} \int_{d_{\min}/d_R}^{d_{\max}/d_R} dy K_j(\kappa R_0 + y\kappa d_R) \exp\left(-\frac{y^2}{2}\right)$$
$$+\frac{1}{2} K_j(\kappa(R_0+d_{\min}))\left[1-\mathrm{erf}\left(-\frac{1}{\sqrt{2}}\frac{d_{\min}}{d_R}\right)\right] + \frac{1}{2} K_j(\kappa(R_0+d_{\max}))\left[1-\mathrm{erf}\left(\frac{1}{\sqrt{2}}\frac{d_{\max}}{d_R}\right)\right],$$

(27)

$$g_{img}(n, \kappa R_0, \kappa d_R, d_{max}/d_R, d_{min}/d_R; a) = \frac{1}{\sqrt{2\pi}} \sum_{j=-\infty}^{\infty} \int_{d_{min}/d_R}^{d_{max}/d_R} dy K_{n-j}(\kappa R_0 + y\kappa d_R) K_{n-j}(\kappa R_0 + y\kappa d_R) \frac{I'_j(\kappa a)}{K'_j(\kappa a)}$$

$$\exp\left(-\frac{y^2}{2}\right) + \frac{1}{2} \sum_{j=-\infty}^{\infty} K_{n-j}(\kappa(R_0 + d_{min})) K_{n-j}(\kappa(R_0 + d_{min})) \frac{I'_j(\kappa a)}{K'_j(\kappa a)} \left[1 - \text{erf}\left(-\frac{1}{\sqrt{2}} \frac{d_{min}}{d_R}\right)\right]$$

$$+ \frac{1}{2} \sum_{j=-\infty}^{\infty} K_{n-j}(\kappa(R_0 + d_{max})) K_{n-j}(\kappa(R_0 + d_{max})) \frac{I'_j(\kappa a)}{K'_j(\kappa a)} \left[1 - \text{erf}\left(\frac{1}{\sqrt{2}} \frac{d_{max}}{d_R}\right)\right].$$

(28)

As well as replacing the bending and electrostatic terms with their averages, there also a couple of other important differences between Eqs. (19) and (23). One is that the third term in Eq. (23), the contribution from steric interactions now depends on $d_R$, as well as $R_0 - 2a$. Another is the appearance two entropic terms, the fourth and fifth terms, that depend on $\theta_R$ and $d_R$, which want to maximise the latter quantity, while steric, bending and electrostatic terms want to restrict its value.

Equations on $\theta_R$, $d_R$, $R_0$ and $\eta_0$ are now got through the conditions

$$\frac{\partial f_{Braid}}{\partial \theta_R} = 0, \quad \frac{\partial f_{Braid}}{\partial d_R} = 0, \quad \frac{\partial f_{Braid}}{\partial \eta_0} = 0 \text{ and } \frac{\partial f_{Braid}}{\partial R_0} = 0. \tag{29}$$

The first condition in Eq. (29) we'll look at in detail, the other equations can be generated by combining Eqs. (23) and (29), and general forms for them can be found in Ref. [41]. From this first condition we obtain the equation

$$0 = -\frac{1}{2l_p \theta_R^3} + \frac{l_p \theta_R^3}{d_R^2} - \frac{2\theta_R l_p}{R_0^2} \tilde{f}_1(R_0, d_R, R_0 - 2a, -(R_0 - 2a)) \sin^2\left(\frac{\eta_0}{2}\right) - \frac{M\theta_R}{2k_B T} \frac{\tilde{f}_2(R_0, d_R, R_0 - 2a, -(R_0 - 2a))}{R_0} \sin\left(\frac{\eta_0}{2}\right)^{-1}.$$

(30)

If we neglect the last two terms in Eq. (30), the contributions from the bending elastic energy and the work term we simply recover an old result (see Ref. [38]),

$$\theta_R^2 = \frac{1}{2^{1/3}} \left(\frac{d_R}{l_p}\right)^{2/3}. \tag{31}$$

An important point to realize is that if we were to substitute Eq. (31) into Eq. (23) and replace the averages of the bending energy terms and electrostatic energy terms with their unaveraged values calculated at $R_0$, on minimization with respect to $d_R$, we would recover back Eq. (19) for the free energy prior to minimization over $R_0$ and $\eta_0$. However, it is far more physically appropriate to consider the thermal averages of the bending energy and electrostatic interaction energy in the free energy.

In general we find that solution to Eq. (30) is well approximated by the formula

$$\theta_R \approx \left(\frac{d_R}{l_p}\right)^{1/3} \frac{1}{\left(4\Gamma^2 + \frac{32}{9}\Gamma + 2^{4/3}\right)^{1/8}}, \qquad (32)$$

where

$$\Gamma = \left(\frac{d_R}{l_p}\right)^{4/3} \left(\frac{Ml_p}{2R_0 k_B T} \tilde{f}_2(R_0, d_R, R_0 - 2a, -(R_0 - 2a))\sin\left(\frac{\eta_0}{2}\right)^{-1} - \frac{2l_p^2}{R_0^2} \tilde{f}_1(R_0, d_R, R_0 - 2a, -(R_0 - 2a))\sin^2\left(\frac{\eta_0}{2}\right)\right). \qquad (33)$$

## 3. Results

### *3.1 Comparing Self consistent treatment against simple model*

Now, we compare the results obtained from Eqs. (19) and (21) with those obtained from Eqs. (23), (29) and (32). We examine the differences in $\eta_0$, $R_0$, $M$ and $z_T$ as functions of $N$ (the number of braid pitches) between the two approximations, for two pulling force values of $F = 2\,\text{pN}$ and $8\,\text{pN}$ and Debye screening lengths $\lambda_D = 30.99\,\text{Å}$ and $9.8\,\text{Å}$, which roughly correspond to 1:1 monovalent salt concentrations of $10\,\text{mM}$ and $100\,\text{mM}$, respectively.

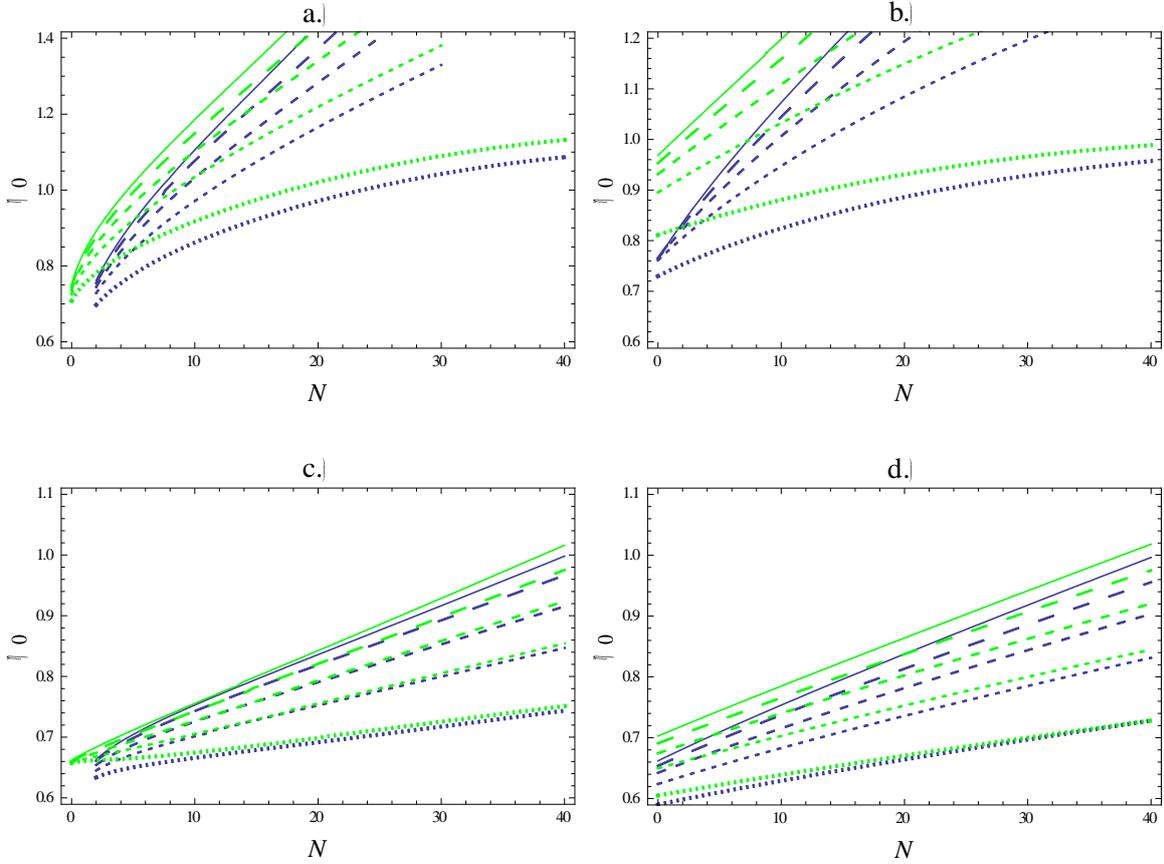

Fig.2. Graphs comparing the tilt angles calculated from the self-consistent treatment with those from the simple model. In all plots, the green (light) curves are generated using the simple model, while the blue (dark) curves are generated using the self-consistent treatment. In Figs. a.) and b.) we use a Debye screening length of $\lambda_D = 30.99\text{Å}$, while in Figs. c.) and d.) $\lambda_D = 9.8\text{Å}$ is used. A pulling force of $F = 1\text{pN}$ in Figs. a.) and c.), and in Figs. b.) and d.) a pulling force of $F = 8\text{pN}$, is used. The solid, long dashed, medium dashed, short dashed and dotted dashed lines correspond to values of $\theta = 0, 0.2, 0.4, 0.6, 0.8$, respectively.

In Fig.2 we present plots for the average tilt angle. We see, generally, that the self-consistent approximation (using Eq. (23)) has a lower value of $\eta_0$ than the results determined from Eq. (19). This difference between the two approximations is most pronounced when $\lambda_D = 30.99\text{Å}$. Also, the difference between the two approximations increases with the increase in the magnitude of the repulsive electrostatic interaction with decreasing $\theta$. The difference can be accounted in the following way. Averaging the bending energy terms, in the self-consistent treatment, as opposed simply calculating them at $R_0$ enhances these terms, favouring a smaller value of $\eta_0$. The size of this enhancement is affected by $d_R$, which is in turn affected by $R_0$, as both electrostatic terms and bending terms help to determine $d_R$, as well as the steric interaction. As one reduces the value of $\theta$ and increases $\lambda_D$, one increases the value $R_0$, since one increases the electrostatic repulsion (see Fig. 3 below). This increase in $R_0$ has the tendency to increase $d_R$. Therefore, the difference is most pronounced for small $\theta$ and large $\lambda_D$. Also, we see that in all cases considered, reducing the value of $\theta$ and increasing $\lambda_D$ both increase $\eta_0$. This is again attributable to the increase in $R_0$; here it weakens the bending energy terms in favour of the moment terms, which causes such an increase in $\eta_0$.

In Fig. 3 we present plots of $R_0$, as a function of the number of turns, for the two approximations. As we decrease $\theta$ and increase $\lambda_D$, we increase the amount of electrostatic repulsion in our system, which pushes up $R_0$. By increasing the pulling force $F$ we reduce $R_0$ in the braid. This is because one requires a larger value of $M$ to stabilize a braid; this larger value forcing the two molecules closer together. We find that $R_0$ is always larger for the self-consistent approximation than for the simpler approximation. One reason for this increase, when we include undulations about $R_0$ into the electrostatic energy, is that this increases the amount of repulsion by enhancement of these terms. A second reason is an increase in the amount of repulsion due to entropy loss when confining the molecules to the braid, due to a reduction in $\langle \delta R(s)^2 \rangle$ when it is self consistently calculated. The difference in $R_0$ between the two approximations is most pronounced at $\lambda_D = 9.81$. The explanation for this is that undulations enhance the electrostatics much more at $\lambda_D = 9.81$ than at $\lambda_D = 30.99$. This is due to the fractional increase in electrostatic energy from reducing $R$ being much larger for the former case, so that the undulations about $R_0$ that reduce $R$ strengthen the electrostatic interactions more here. Also, this difference between the two approximations is most pronounced at $F=2\,\text{pN}$ and for small values of $N$, this is because at these values $d_R$ is largest.

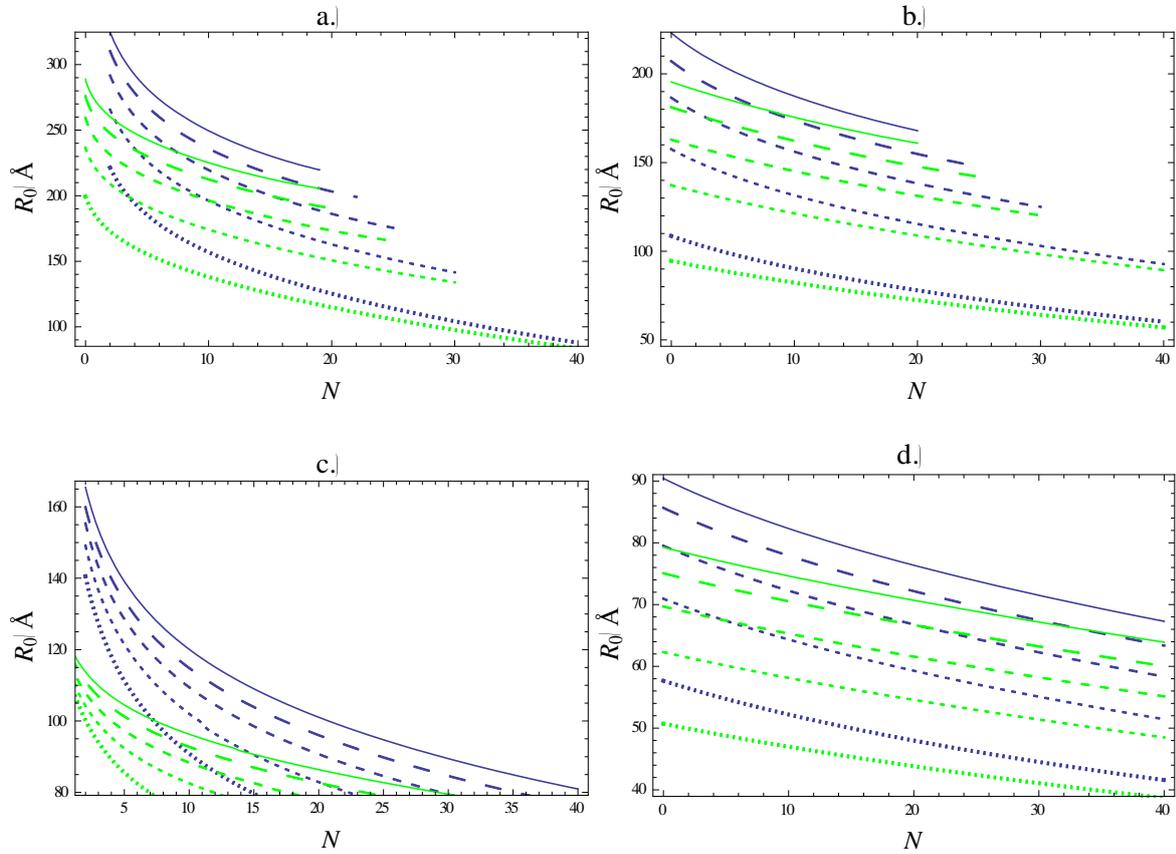

Fig. 3. Graphs comparing the braid radius calculated from the self-consistent treatment with the simple model. In all plots, the green (light) curves are generated using the simple model, while the blue (dark) curves are generated using the self-consistent treatment. In Figs. a.) and b.) a Debye screening length of $\lambda_D = 30.99\,\text{Å}$ is used, while in Figs. c.) and d.) $\lambda_D = 9.8\,\text{Å}$ is used. A pulling force of $F = 2\,\text{pN}$ in Figs. a.) and c.), and in Figs. b.) and d.) a pulling force of $F = 8\,\text{pN}$, is used. The solid, long dashed, medium dashed, short dashed and dotted dashed lines correspond to values of $\theta = 0, 0.2, 0.4, 0.6, 0.8$, respectively. Note that in a.) and b.) some of the curves have been terminated, as $\eta_{end}$ has become too large for the model to be valid.

In Fig. 4 we examine the moment $M$ as a function of the number of braid turns. The tendency, here, is for the self-consistent approximation to give a slightly larger value for the moment than for the simple approximation. At the low force value of $F = 2\,\text{pN}$ this difference is most apparent, especially for $\lambda_D = 9.8\,\text{Å}$. This difference is due to the increase in electrostatic repulsion from the effect of averaging the electrostatic energy over the braid undulations, thereby increasing the amount of moment needed to do work against repulsive forces. These forces need to be overcome to bring the molecules close together, producing a braid of $N$ turns.

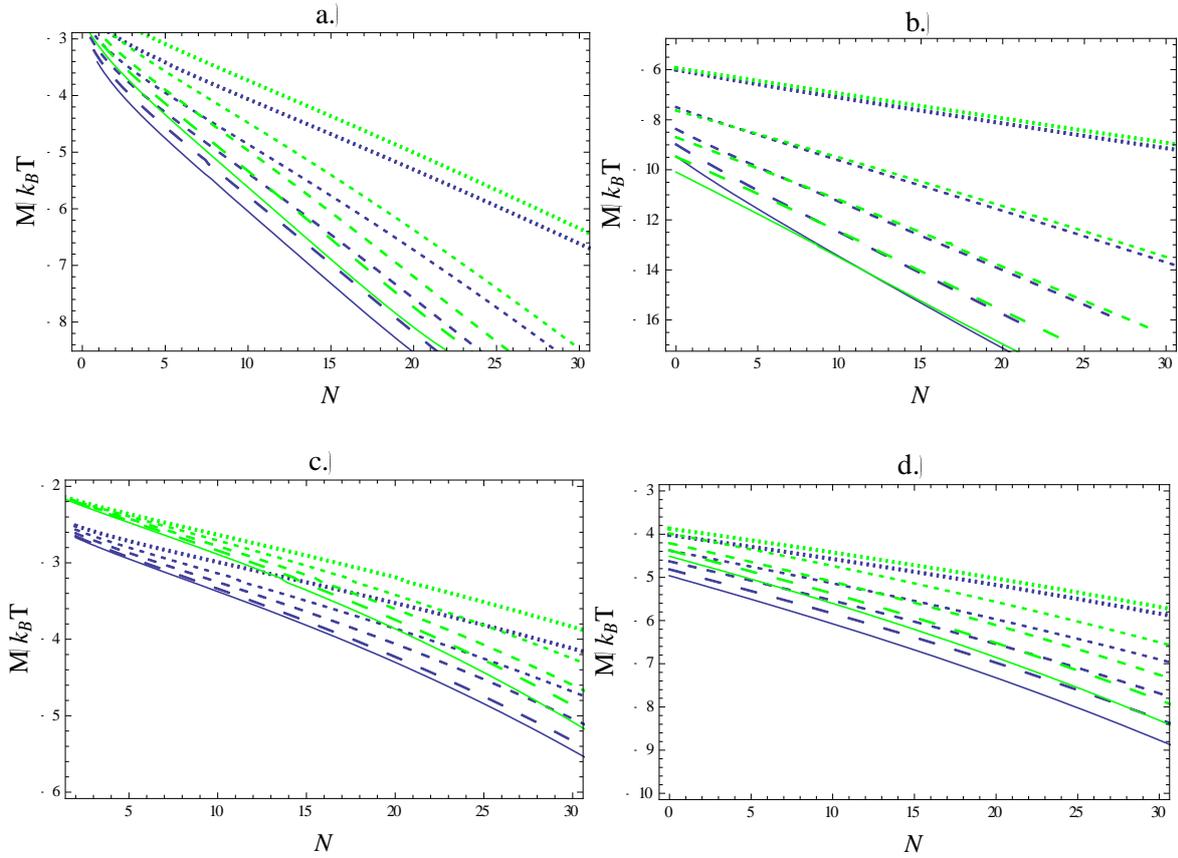

Fig.4. Graphs are shown comparing the relationship between the applied moment $M$ and the number of braid turns, calculated from the self-consistent treatment with that from the simple model. In the calculations, the values $L = 36000\text{Å}$ and $b = 12000\text{Å}$ are used. The same colour coding is used as in the previous figure. In Figs. a.) and b.) a Debye screening length of $\lambda_D = 30.99\text{Å}$ is used, while in Figs. c.) and d.) $\lambda_D = 9.8\text{Å}$ is used. A pulling force of $F = 2\,\text{pN}$ in Figs. a.) and c.), and in Figs. b.) and d.) a pulling force of $F = 8\,\text{pN}$, is used. The solid, long dashed, medium dashed, short dashed and dotted dashed lines correspond to values of $\theta = 0, 0.2, 0.4, 0.6, 0.8$, respectively. Note that all the moment curves have the symmetry property $M(N) = -M(-N)$.

Last of all, we compare the extensions in Fig. 5, or end to end distance, $z_T$. In most plots, the self-consistent approximation gives a slightly larger value of $z_T$ for fixed $N$. We might have expected the opposite (Ref. [33]), as for the self-consistent approximation we obtain a larger values of $R$ at fixed moment $M$, which would certainly be the case if $\eta_0$ and $L_b$ remained fixed. However, the tilt angles $\eta_0$ are smaller for the self-consistent approximation than the simple approximation (see Fig.2) as well as $L_b$. Also, we see that the differences between two approximations at $\lambda_D = 9.8\text{Å}$ seem to be very slight. Again, these differences, in most cases, are reduced with increasing $\theta$, as we would expect.

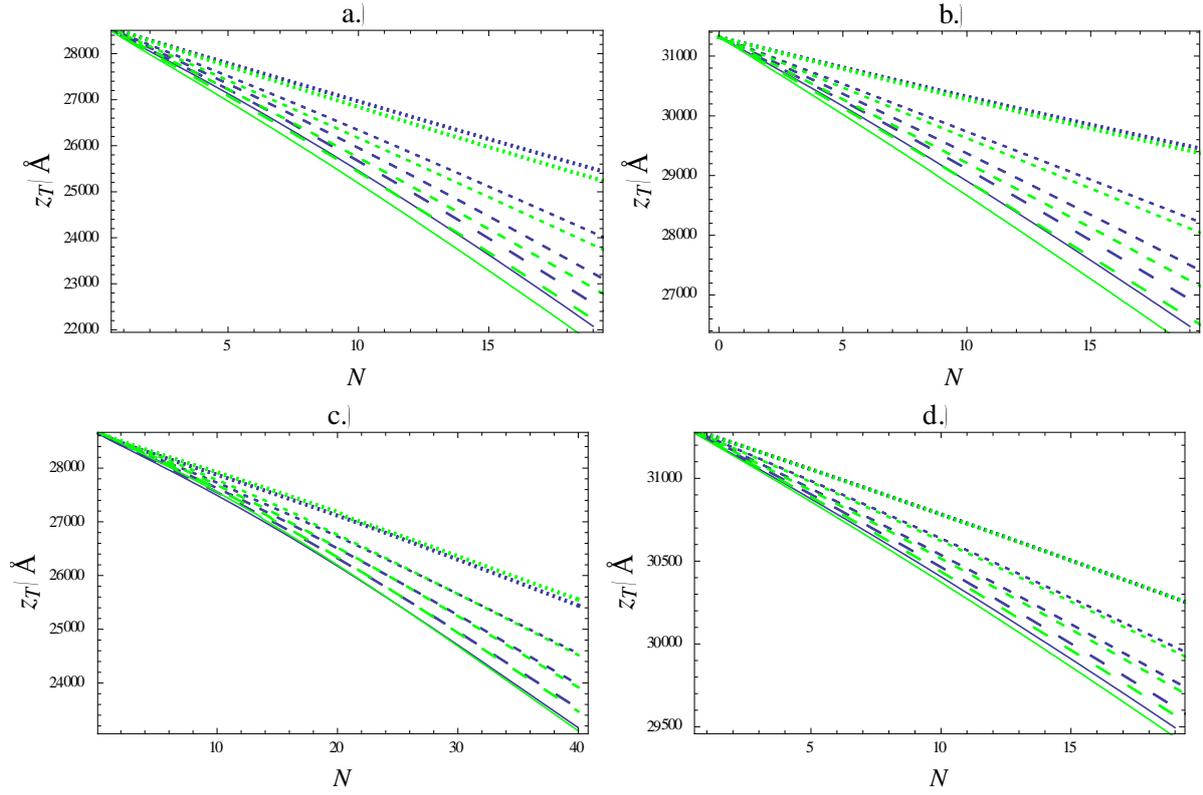

Fig. 5. Graphs are shown comparing the extension $z_T$ as a function of the number of braid turns calculated from the self-consistent treatment with the simple model. In the calculations, the values $L = 36000 \text{Å}$ and $b = 12000 \text{Å}$ are used. The same colour coding is used as in the previous figure. In Figs. a.) and b.) a Debye screening length of $\lambda = 30.99 \text{Å}$, while in Figs. c.) and d.) $\lambda = 9.8 \text{Å}$ is used. In Figs. a.) and c.) a pulling force of $F = 1 \text{pN}$, and in Figs. b.) and d.) a pulling force of $F = 8 \text{pN}$, is used. The solid, long dashed, medium dashed, short dashed and dotted dashed lines correspond to values of $\theta = 0, 0.2, 0.4, 0.6, 0.8$, respectively. Note that the extension curves have the symmetry property $z_T(N) = z_T(-N)$.

## 3.2 Comparing Self consistent treatment against experimental data

We now match the self-consistent approximation with the experimental extension curves of Ref. [16]. We have essentially two fitting parameters $b$ and $\theta$, the latter is a fitting parameter as it can only be determined from the extension data; it cannot be measured independently. Also, we should point out that the value of $b$ cannot be controlled precisely in the experimental set up of Ref. [16], and may vary from one experiment to the next. One can fit $b$ using only the extension data between $n = -1/2$ and $1/2$ where there is no braid. However, note as was stated in Ref. [16] that there is a $10\%$ error in the values of the measured applied pulling force $F$. Therefore, because of this and the fact that we do not have in all cases the available data we fit $b$ globally to the extension curves. Also due to this uncertainty in $F$ we have used the generic value of $l_p \approx 500 \text{Å}$ as opposed to a fine tuned value. It is worth pointing out that $l_p$ does vary slightly with salt concentrations between $10 \text{mM}$ and $100 \text{mM}$ [45].

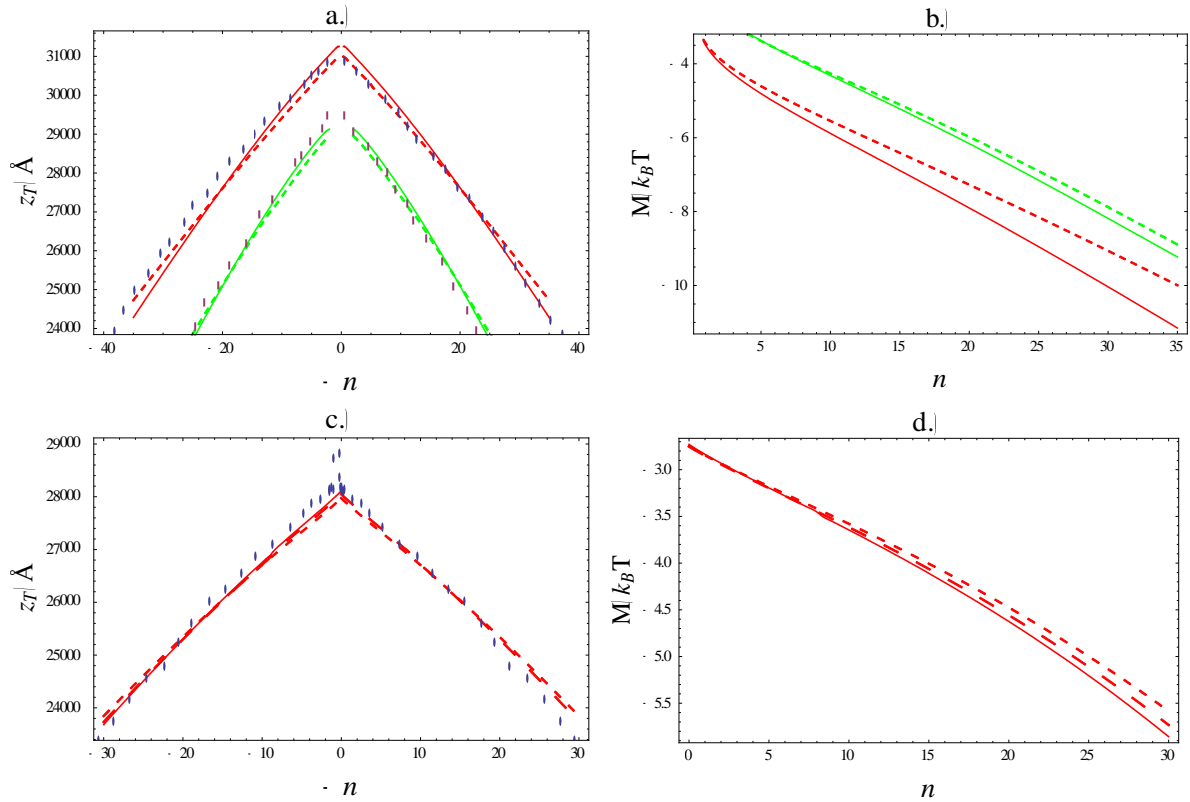

Fig. 6. This figure shows the fits to the experimental extension curves along with the predicted applied moment for those fits. In Fig. a.) theoretical extension curves are fitted to experimental data from Ref. [16] at a monovalent salt concentration of $10\,\mathrm{mM}$. The circles are experimental data for which a pulling force of $F = 4\,\mathrm{pN}$ was used, whereas the squares correspond to a pulling force of $F = 2\,\mathrm{pN}$. The red (top) curves are theoretical curves calculated from the self-consistent approach at a pulling force of $F = 4\,\mathrm{pN}$, and the green (bottom) curves are theoretical curves calculated using $F = 2\,\mathrm{pN}$. For the solid lines a value of $\theta = 0.5$ was used and for the dashed curve a value of $\theta = 0.6$ was used. Fig b.) shows the predicted applied moment for the fits to the $10\,\mathrm{mM}$ extension curve data, as a function of the number of turns of the bead, for the values $\theta = 0.5$ (shown by solid line) and $\theta = 0.6$ (shown by dashed line) at the force values $F = 2\,\mathrm{pN}$ (higher curves, green) and $F = 4\,\mathrm{pN}$ (lower curves, red). Note that the moment curves have the symmetry property $M(n) = -M(-n)$. In Fig c.) we show fits of the experimental data of Ref. [16] at monovalent salt concentration $100\,\mathrm{mM}$ with pulling force $F = 2\,\mathrm{pN}$. The experimental data is given by blue circles, while the theoretical curves are red; the solid, long dashed and medium dashed lines correspond to $\theta = 0.2, 0.3$ and $\theta = 0.4$ respectively. In Fig d.) we show the predicted applied moment curves for the fits to the $100\,\mathrm{mM}$ extension data. Again, the solid, long dashed and medium dashed lines correspond to $\theta = 0.2, 0.3$ and $\theta = 0.4$, respectively.

To roughly quantify the goodness of the fit, we may compute a normalized variance for $N_s$ experimental data points that lie in the region where the theoretical data curves are valid (i.e. $\eta_{end} \lesssim 1.8$ and $|n| > 1/2$). This is defined as

$$\sigma^2 = \sum_{j=1}^{N_s} \left[ \frac{(z_T(n_j) - z_j)}{z_T(n_j)} \right]^2 \tag{34}$$

where, for experimental each experimental data point, we have the coordinates $z_j$ and $n_j$ for the extension and the number of bead turns. In Eq. (34), the theoretical curve is given by an interpolation function $z_T(n_j)$ generated from the numerical data. To obtain the best fit for $b$, we changed $b$ in steps of $100\,\text{Å}$ and computed $\sigma^2$ for each of the theoretical curves. For each value of $\theta$, the value of $b$ that generates $z_T(n)$ with the smallest value of $\sigma^2$ was judged to be the best fit.

We see that we can obtain good fits to the experimental data of Ref. [16] in Fig. 6. As we saw in Fig. 5 there may only be a slight difference in the extension curves between the two approximations. The improvement over the preliminary fits of Ref. [33], may in fact be mostly attributable to a slight difference in the expression for $\alpha_\eta$ and the additional term $\propto M^2$ in the free energy due to taking account of $-2\pi MWr_b$ in Eq. (16), the work term. However, there are significant differences in $\eta_0$, $R_0$ and $M$ between the two approximations, and the self-consistent approximation reflects better physics

Unfortunately, quite a large range of values of $\theta$ fit the extension curves for $100\,\text{mM}$ and $10\,\text{mM}$, though with quite different fitted values of $b$ (see Table 1 and 2). If the value of $b$ was fixed, we would see the difference that is seen in Fig. 5, but some of this difference is offset by adjusting $b$. We find that for $10\,\text{mM}$ the values $\theta = 0.5$ and $\theta = 0.6$ fit the data well (see Fig.6), with the values of $b$ given in Table 1. The fits for $\theta = 0.4$ and $\theta = 0.7$ are significantly worse. For $100\,\text{mM}$, we find that $\theta = 0.2, 0.3$ and $0.4$ fit the data well; the best fit being $\theta = 0.2$. Again, all of these give different values of $b$ (see Table 2) and different curves for $M$ as a function of $n$ (see Fig. 6). However, we have refrained from going to $\theta = 0.1$, as we think this represents a rather unrealistic value of the charge compensation. All theoretical curves are terminated when roughly $\eta_{end} \approx 1.8$, as at this point buckling of the braid may have already occurred and the theory is not really strictly valid when $\eta_{end} > \pi/2$, although one can probably extrapolate slightly to our chosen value. The variances for the best fits to the $10\,\text{mM}$ and $100\,\text{mM}$ monovalent salt concentration data are given in Tables 1 and 2, respectively.

| Charge compensation $\theta$ | Fitted value of b/Å | | Normalized variance Squared of best fit | |
|---|---|---|---|---|
| | $F = 2\,\text{pN}$ | $F = 4\,\text{pN}$ | $F = 2\,\text{pN}$ | $F = 4\,\text{pN}$ |
| 0.4 | 8400 | 7300 | 0.00102 | 0.00438 |
| 0.5 | 9100 | 8600 | 0.00070 | 0.00308 |
| *0.6* | *9900* | *9100* | *0.00223* | *0.00264* |
| 0.7 | 10900 | 10800 | 0.00769 | 0.01399 |

Table 1. This table shows the fitted values of the distance $b$ between the two sets of DNA ends (varied in steps of $100\,\text{Å}$) and normalized variance for the $10\,\text{mM}$ monovalent salt data. The latter is calculated with Eq. (34),

as a measure of how well each choice of $\theta$ fits the experimental data. Shown in the table are fits for the two pulling force values of $F = 2\,\text{pN}$ and $F = 4\,\text{pN}$.

| Charge compensation $\theta$ | Fitted value of b/Å $F = 2\,\text{pN}$ | Normalized Variance Squared $F = 2\,\text{pN}$ |
|---|---|---|
| 0.2 | 13600 | 0.00109 |
| 0.3 | 13800 | 0.00162 |
| 0.4 | 13900 | 0.00270 |
| 0.5 | 14100 | 0.00407 |

Table 2. This table shows the fitted values of $b$ the distance between the two sets of DNA ends (varied in steps of $100\,\text{Å}$) and normalized variance for the $100\,\text{mM}$ monovalent salt data . The latter is calculated with Eq. (34), as a measure of how well each choice of $\theta$ fits the experimental data. Shown in the table are fits for the pulling force value of $F = 2\,\text{pN}$.

## 4. Discussion and outlook

In the results section, we started by comparing the self-consistent determination of the mean-squared amplitude of fluctuations with a cruder, but simpler, approach that was used in Refs. [33] and [34], also implied in the calculations of Refs. [17,25,26,27]. This self-consistent calculation is akin to the approach used by Ref. [24] to describe the statistical mechanics of braiding, which was used successfully to match single molecule twisting data [28]. We found that there is a significant difference between the self-consistent approximation and the simpler approximation for $\eta_0$, $R_0$ and $M$, as functions of the number of braid turns, that grows with the increasing strength of the electrostatic interaction. Though surprisingly, the difference between the two sets of extension curves is slight. Nevertheless, we would still advocate, unless the electrostatic interaction is particularly weak, that the self-consistent approximation is the better one to use, and it contains better physics.

To test this improved theory in describing the braiding of two molecules, we have fitted it against the experimental data of Ref. [16], and have obtained good fits. However, we have not attempted to fit the force values $F = 0.5\,\text{pN}$ and $1\,\text{pN}$, as some of the approximations presented here are not quite valid for such low forces. Indeed, the expressions that are used to determine $\eta_{end}$ ( notably Eq. (22)) and Eqs. (2) and (3) are only valid at sufficiently large pulling force. Though, it is quite possible to extend the theory to these force values by numerically determining $\eta_{end}(F)$, $L_b(F)$ and $f_{wlc}(b,F)$ through the WLC model, but this has yet to be attempted. What is encouraging, in the current fits, is that the values of $\theta$ that fit the $10\,\text{mM}$ data are larger than those that fit the $100\,\text{mM}$ data. This is entirely consistent with the trend suggested by conventional counter-ion condensation theories, involving the solution of the non-linear Poisson-Boltzmann equation, where the charge compensation should decrease from its Manning value [40], at infinite dilution, with increasing salt concentration.

What is quite interesting is that there is a slight asymmetry in the experimental data, particularly seen for a $10\,\text{mM}$ monovalent salt concentration, at a pulling force of $F = 4\,\text{pN}$ (c.f. Fig. 6). This slight asymmetry might be explained by weak correlations between the two helix structures of the braided part of the molecules. At any position along the braid, for a particular configuration of the molecules, we may write the following form for the interaction energy

$$E_{\text{int}} = \int_{-L_b/2}^{L_b/2} ds \left( \mathcal{E}_{dir}(R(s)) + \mathcal{E}_{img}(R(s)) \right) + \int_{-L_b/2}^{L_b/2} ds \left( \mathcal{A}_{1,dir}(R(s)) + \mathcal{B}_{1,dir}(R(s)) \sin \eta(s) \right) \cos\left( \xi_1(s) - \xi_2(s) \right)$$
$$+ \int_{-L_b/2}^{L_b/2} ds \left( \mathcal{A}_{2,dir}(R(s)) + \mathcal{B}_{2,dir}(R(s)) \sin \eta(s) \right) \cos\left( 2(\xi_1(s) - \xi_2(s)) \right),$$

(35)

where $\xi_1(s) = \hat{\mathbf{d}}_1(s) \cdot \hat{\mathbf{v}}_1(s)$ and $\xi_2(s) = \hat{\mathbf{d}}_2(s) \cdot \hat{\mathbf{v}}_2(s)$ are the azimuthal orientations of the minor grooves. The vectors $\hat{\mathbf{v}}_1(s)$ and $\hat{\mathbf{v}}_2(s)$ are perpendicular to $\hat{\mathbf{t}}_1(s)$ and $\hat{\mathbf{t}}_2(s)$, lying along lines connecting the molecular centre lines with the minor grooves (shown in Fig. 1). The terms $\mathcal{A}_{\mu,dir}(R)$ and $\mathcal{B}_{\mu,dir}(R)$ are contributions to the direct electrostatic interaction, due to the helical structure of the molecules; the latter being terms that generate an internal chiral torque [33]. If the helix dependent second and third terms in Eq. (35) are sufficiently large, a preferred average azimuthal alignment $\langle \xi_1(s) - \xi_2(s) \rangle$ is maintained along the braid, and then the strong chiral regime discussed in Ref. [33] holds. However, if the terms are not quite large enough, $\langle \xi_1(s) - \xi_2(s) \rangle$ does not exist in the limit $L_b \to \infty$; there is no preferred average azimuthal orientation between the two grooves. Nevertheless, a term proportional in the free energy to $\sin \eta(s)$ (a chiral torque) is not completely washed out by thermal fluctuations in this state, as was originally suggested in Ref. [33]. While writing this paper, we realized that there is a possibility for weak transient correlations between $\xi_1(s)$ and $\xi_2(s)$ to occur in patches along the molecules, changing as the molecules thermally fluctuate, thereby causing a weak chiral torque. To calculate this weak chiral torque requires a different approach from the strong chiral interaction regime. In this new approach, the second and third terms in Eq. (35) should be treated as perturbations, when calculating the free energy. Such a perturbation approach was considered previously for DNA assemblies [46,47]. The leading order term of the perturbation expansion will still provide Eq. (23) (or Eq. (19)), but there should be a small correction to it from the perturbation expansion that breaks the $n \to -n$ symmetry. Indeed, we hope to look at this correction to the free energy, perhaps, in a later work to see whether it can account for the observed asymmetry. On the experimental side, if this is indeed the explanation for what is seen, we would expect the asymmetry becomes larger on increasing the force, as this brings the molecules closer, as well as by increasing the valance of the counter-ions, which should increase the relative strength of helix specific forces.

We still have yet to include buckling of the braid into the theory; when $|n|$ is sufficiently large, we would indeed expect it. In Fig. 6, we see a slight dip seen in at both $n \approx 20$ and $n \approx -20$ in the $100\,\text{mM}$ extension curve data. If this was experimental error, we would not expect that this dip would occur on both sides of the extension curve of roughly the same value of $n$, perhaps suggesting a different explanation for this feature. Indeed, it is well known in single molecule twisting experiments [11,12,13] that on forming the end loop of a plectoneme the extension of the molecule drops in a discontinuous fashion. Interestingly, no such dip is seen in the $10\,\mathrm{mM}$ data of Ref. [16] (see Fig. 6). This actually is line with theoretical [26,29] and experimental [11,12,13] trends for single molecule twisting; the size of the extension drop reduces and goes away with decreasing salt concentration. Therefore, perhaps this feature is indeed the hallmark of the formation of end loop of a super plectoneme structure, in which the braid axis traces out a plectoneme. Though, it is also quite conceivable that it might be some other type of buckling, if it is not an experimental artefact. It would be interesting to see what a theory incorporating different buckled states would predict as the buckling transition and the type of buckling.

Recently, state of the art braiding experiments have been developed [48] using four optically trapped beads, which potentially offer much greater control over the geometry of the two molecules than those of Ref. [16], most notably $b$. Such experiments, perhaps, offer an opportunity to investigate DNA friction, the recent topic of a preliminary theoretical investigation [49]. In Ref. [49], the effect of the braid geometry was not taken into account; nevertheless it could be built upon using a similar framework to that suggested in Refs. [33] and [34], and this current work. However, a notable technical problem with these experiments is that, at present, only a few braid turns can be accommodated [50]. Therefore, the current model will need to be modified to the regime of a short braid to describe such experiments.

At present, we are working on the possibility of the collapse of the braid into a tighter braided structure. Such a collapse may occur when there is a significant attractive component to the interaction between the two molecules. This might be caused by non-chiral attractive forces or forces dependent on helix structure. We hope to investigate both possibilities. For the latter, collapse into the tightly braided state happens predominantly in left handed braids. Such a possibility has already been investigated and discussed in Ref. [33] in the strong chiral regime, but the extension curves, here, were calculated in the absence of molecular undulations in the braided section. However, we can now incorporate braid undulations [41]. We will also include an estimate of $\sin^2 \eta(s)$ terms in the interaction energy, based on geometric arguments. From these arguments, the helical geometry of the DNA should actually limit the optimum value of the tilt angle $\eta(s)$, even in the absence of a bending rigidity term. In incorporating these two effects, we will see how the collapse of the braid for such forces is qualitatively changed from that of Ref. [33].

## Acknowledgements

D.J. (O') Lee would like to acknowledge useful discussions with R. Cortini , G. King, A. Korte, A. A. Korynshev, E. L. Starostin , G.H.M. van der Heijden and G.J.L. Wuite. This work was initially inspired by joint work that has been supported by the United Kingdom Engineering and Physical Sciences Research Council (grant EP/H004319/1). He would also like to acknowledge the support of the Human Frontiers Science Program (grant RGP0049/2010-C102).

[43] It is worth mentioning that in the Gaussian chain model of two entwined directed polymers, the problem of steric interactions have been solved exactly [F. Ferrari, V. G. Rostiahvili, T. A. Viligis, Phys. Rev. E **71**, 061802], assuming that the polymer diameter is negligible. Also, using a cylinder model for steric interactions an exact solution was obtained for the Gaussian model [R. Podgornik and V. A. Parsegian, Macromolecules **23**, 2265 (1990)]. However for the wormlike chain, things are rather more difficult due to the second derivative. Here, the resulting Schrodinger like equation for the probability amplitude depends on both the direction of the tangent vector and position of the molecular chain [H. J. Kleinert, J. Math. Phys. **27**, 3003 (1986)], and is generally difficult to solve.

[44] Strictly speaking from a mathematical perspective, this integral for the computation of writhe, as well as the definition of linking number, should be for a closed loop. However, on can think of a phantom chain that

contributes to neither $Lk_b$ nor $Wr_b$ that connects the two ends of the braid to form a closed loop. Further justification of this extension to such non closed loop systems can be found in [E. L. Starostin, *Physical and Numerical Models in Knot Theory Including Applications to the Life Sciences* (eds J.A. Calvo, K.C. Millett, E.J. Rawdon and A. Stasiak) (World Scientific, Singapore, 2005), Chapter 26, pp. 525-545]